\documentclass[reprint,showpacs,amsmath,amssymb,pra,aps]{revtex4-1}
\usepackage{graphicx}% Include figure files
\usepackage{dcolumn}% Align table columns on decimal point
\usepackage{bm}% bold math
\usepackage{natbib}
\begin{document}

%\preprint{APS/123-QED}

\title{The relation between optical instabilities and absorbed material in photoluminescence with [0001] InGaN single quantum well.}% Force line breaks with \\

\author{T. Tsutsumi}
\author{R. Micheletto}
\affiliation{Department of Nanosystem Sciences, Yokohama City University, Yokohama 236-0027, Japan}

\author{G. Alfieri}
\altaffiliation[Now at ]{ABB Corporate Research Center, Switzerland}
\affiliation{Department of Electronic Science and Engineering, Kyoto University, Kyotodaigaku-katsura, Nishikyo, Kyoto 615-8510, Japan}

\author{Y. Kawakami}
\affiliation{Department of Electronic Science and Engineering, Kyoto University, Kyotodaigaku-katsura, Nishikyo, Kyoto 615-8510, Japan}

\date{\today}

\begin{abstract}
%III-V nitrides, like InGaN, are attractive semiconductors for light-emitting diodes (LED) because their emission ranges from the near ultraviolet to near infrared. 
In this letter, we aim to elucidate the physical mechanism of the so called optical memory effect and blinking phenomenon observed in InGaN single quantum wells (SQW).
We have found that the optical response of both memory effect and blinking phenomenon, is affected by different excitation wavelengths and by the change of gas adsorption on the crystal surface. A model that reproduce dynamics of the coverage of absorbed gas molecules on the sample surface is given and compared with experimental data with evident match. \end{abstract}
\pacs{Valid PACS appear here}% PACS, the Physics and Astronomy

\maketitle
\begin{figure}[!ht]
\begin{center}\leavevmode
\includegraphics[width=7cm]{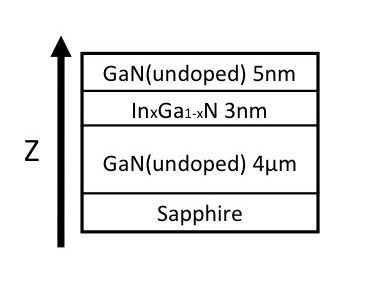}
\caption{The structure of our sample. A 4 mm sapphire substrate is at the base of a layer of undoped GaN (4 $\mu$ m), an active layer of InGaN 3 nm thick and a final 5 nm capping layer of undoped GaN.}
\end{center}
\label{fig:1}
\end{figure}
\begin{figure}[!ht]
\begin{center}\leavevmode
\includegraphics[width=8.5cm]{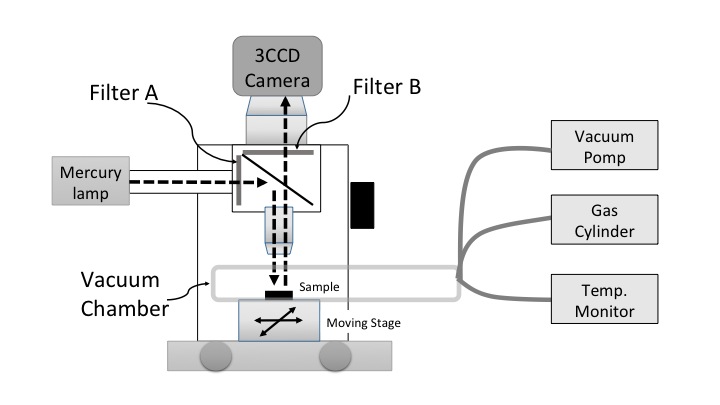}
\caption{The scheme of out selective excitation microscope. A mercury lamp is filtered by the filter A to select the excitation wavelength ($\lambda$=405 nm or $\lambda$=365 nm). A second colored filter in B is used to cut away the excitation signal and collect only the photoluminescence (the filter is centered about $\lambda$=500nm). Sample is held in a termally controlled holder enclosed in a vacuum chamber connected to various gas cylinders.}
\label{fig:scheme}
\end{center}
\end{figure}
\begin{figure}[!ht]
\begin{center}\leavevmode
\includegraphics[width=8.5cm]{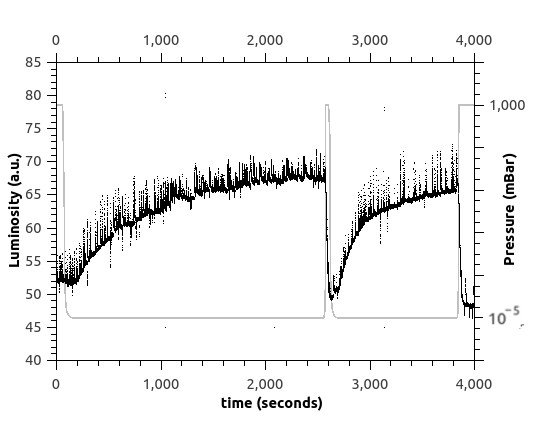}
\caption{The luminosity time dependence when sample is subject to pressure changes. The gray line represent air pressure, that initially is kept at 1000 mBar and then drawn to 10$^{-5}$mBar by a vacuum pump. As soon as the pressure drops, photoluminescence grows up to 150$\%$ the initial value. Excitation wavelength is 365 nm, photoluminescence is filtered by a 500nm color filter and it is centered around 540nm. Same test is repeated twice in this recording. }
\label{fig:photoeff}
\end{center}
\end{figure}
\begin{figure}[!ht]
\begin{center}\leavevmode
\includegraphics[width=8.5cm]{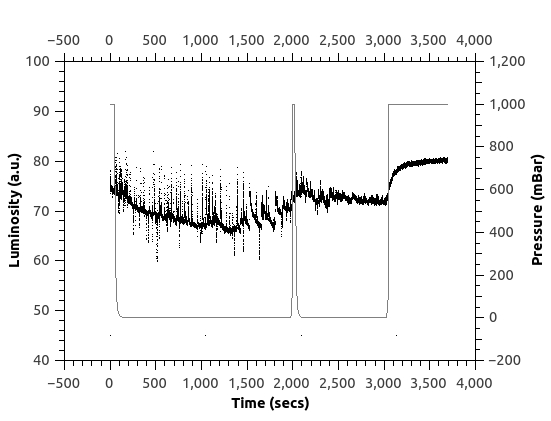}
\caption{The profile of luminosity of a sample excited with 405nm and exposed to variation of pressure in air. The experiment is performed in the same conditions as in figure \ref{fig:photoeff}.}
\label{fig:photoeff405}
\end{center}
\end{figure}
\begin{figure}[!ht]
\begin{center}\leavevmode
\includegraphics[width=8.5cm]{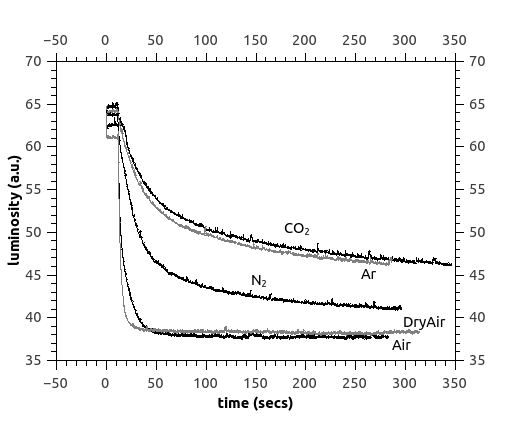}
\caption{ The decay of luminosity when a InGaN sample is brought from high vacuum to atmospheric pressure with different gases. The gases containing Oxygen show the most prominent influence.}
\label{fig:Gas}
\end{center}
\end{figure}

InGaN-based LED of green-ultraviolet region and white LED with phosphor have come into practical use. However, the crystal growth methods and understanding of InGaN devices are still affected by the lack of knowledge about the theoretical background on different phenomena\cite{stri:1992}. For instance,  despite the large concentration of threading dislocations ($10^8-10^{10}~cm^{-2}$)\cite{shik:1997,kane:2006} due to lattice mismatch\cite{zhel:1997}, InGaN optical devices achieve very high internal quantum efficiency and optical emission. In addition, InGaN quantum wells with high Indium composition domains, such as quantum dots (QDs), are known to induce local excitation emission (bright spots/points)\cite{mich:2004,naru:1997}. 
Recently, instability blinking\cite{oika2011,mich:2006a} and optical memory effect\cite{feld2009}], observed by PL in InGaN Single Quantum Well (SQW), were reported and associated to the presence of defects and strain within the crystal\cite{chic:1996}. 
Instability blinking was found in CdSe nanocrystal\cite{nirm1996,yoshi2007} and epitaxial grown film\cite{seuf2001,kauf2013}, ZnCdSe quantum dots (QD)\cite{zhan1998,soto2013}, GaAs QD\cite{bert1999,lyon2010}, InP QD\cite{kuno2001,sugi2002,duan2009} and Porous silicon. In InGaN SQW, instability blinking arises around a QD in a region of few $\mu$m of diameter and the flashing intensity width is temperature dependent\cite{jano2008}. Therefore, the phenomenon was associated to beating of slightly different thermal wave vibration creating unstable optical blinking\cite{mich2013}. 
On the other side, the so called “optical memory effect” was first found in the GaN epitaxial film\cite{feld2009,wall2004}. Optical memory effect in InGaN is the phenomenon in which photoluminescence emission becomes gradually stronger on a time scale of seconds to few minutes and depends from the previous illumination history of the sample (hence the name optical memory effect). 
A complete and universally accepted theory to explain these two phenomena is unknown.\\
%\section{Methods}
In this study, we employed InGaN SQW grown in (0001) direction with the MOVPE method (fig. \ref{fig:1}). 
The InGaN SQW layers are composed of an undoped GaN layer (4 $\mu m$) on a sapphire substrate, an InGaN active layer (3nm), and an undoped GaN layer (5nm). The main peak of the bulk macroscopic photoluminescence was about 540nm. The SQW was optically characterized by the experimental setup shown in Fig. \ref{fig:scheme}. 
The sample in placed in a vacuum chamber (RC102-CFM, CIA, Inc.), connected to a turbo-molecular vacuum pump (TSH 071 E, Pfeiffer), to a temperature controller (Model 32, Cryogenic Control Systems, Inc.) and a gas cylinder.
The dynamics of the photoluminescence is detected by using a selective excitation fluorescence microscope (BX51WX, Olympus) coupled with a CCD camera (HDR-SR1, SONY). The time variation of the light emission surface is recorded as video data at 60 frames/second. Ultraviolet Hg lamp was used as excitation light (365nm and 405nm emission lines). The 365nm light excites both of InGaN layer and GaN layers. On the other hand, 405nm light excites only the InGaN layer.\\  
 In Fig. \ref{fig:photoeff}, we show the effects of air pressure (from 1000 mbar down to $1.0\times10^{-5}$mbar) on the photoluminescence. It can be seen that there is a variation of the overall emission intensity when changing the degree of vacuum. The gray curve represent the change of pressure in log scale, whereas the dots are the sample luminosity, averaged on the entire surface. The sample is excited by a 365 nm signal.
We see that whole emission intensity varies dynamically depending on pressure. In high vacuum, the emission improves and grows to a pleteau 40$\%$ higher than the baseline at one atmosphere. Also, the luminosity distribution becomes homogeneous. On the contrary, at air pressure, the intensity drops to lower values and the emission spatial distribution is less uniform with many bright points, some of those result to be blinking.\\
On the other hand, excitating the sample at 405nm we observe small changes of intensity in response to pressure, the PL emission on the surface is homogeneous and we observe no blinking points (figure \ref{fig:photoeff405}). Since with 405 nm we do not excite the InGaN layer, this suggests that the blinking and those Intense Luminous Centers are associated with the GaN/InGaN interface in proximity of the surface.\\
%In figure \ref{fig:airPr} the decay of luminosity when pressure returns to 100 mBar is shown with an almost perfect single exponential decay fit (continous curve).
To understand the physical mechanism that drives this effect, we have to consider that, in general, adsorption by Van der Waals' forces on the surface of a material changes with the degree of vacuum. In addition, the presence of blinking points and overall emission intensity depends on pressure, suggesting that the phenomenom is driven by a change in the amount of adsorbed species on the surface. 
 In order to prove this, we proceeded by creating vacuum in a room temperature vacuum chamber (about 300 K and $1\times 10^{-5}$ mbar) and then by injecting gas, like air, pure dry air (O$_2$ : N$_2$ = 2 : 8), N$_2$ gas, Ar gas, and CO$_2$ into the chamber, reaching up to a pressure of 1000 mbar (1 atm). Following the injection of each gas, we observed the variation of PL emission. Since there are almost no optical emission variations when InGaN SQW is excited with 405nm wavelength, in the following experiments we will only used 365nm excitation light.
  In Fig. \ref{fig:Gas} it is shown the PL profile when a sample excited by 365nm wavelength is brought back from high vacuum condition to 1000 mbar for each injected gas. It can be seen that air and pure dry air greatly reduce the PL emission. On the other hand, Ar and CO$_2$ gas have a low impact, whereas N$_2$ gas have an intermediate effect. Based on this evidence, we put forward the hypothesis that molecular oxygen is the most effective specie and the main cause of emission variation.
 This is in accordance with what reported by both Zywietz et al~\citep{Zyw}, who showed that GaN surface is very active towards oxygen incorporation, and by Pearon et al.~\cite{Pea}, who experimentally  demonstrated that  oxygen can be found in GaN up to a depth of 180 nm. More recently, density functional theory calculations revealed that oxygen can be easily incorporated into InGaN mono-layer QW (MLQW)~\cite{Alfi2015} in accordance with the experimental results of Kappers et al.~\cite{Kapp2015} who reported that growing N-poor InGaN leads to high levels of oxygen incorporation. The detrimental effects of oxygen in InGaN are known: it can compensate dopants, making the growth of p-type InGaN difficult\cite{miao2013} and it has been held responsible for the degradation of InGaN/GaN LED\cite{mose2011,yang2011,okad2014}.
  %\section{A model for the photoluminescence dynamics}
  We model the relation between emission intensity and ambient pressure in order to give an explanation for the decreased emission at higher pressure. We base ourselves on a simple molecular coverage dynamics theory. We suppose that the optical emission, generated at the InGan/GaN interface, is decreased by the presence of absorbed molecules on the surface accordingly to this linear relation:
\begin{equation}
I(t)=-n(t)\beta+I_{max}
\label{eq:prop}
\end{equation}
where $I_{max}$ is an offset, $\beta$ a coefficient and $n(t)$ the number of molecules present on the surface at time $t$. The term $I_{max}$ represents the maximum luminosity the sample can have if there are no molecules on the surface. 

To model how $n(t)$ varies in time, we consider a space where molecules can be absorbed or desorbed as an uni-dimensional line representing the surface of the sample. For simplicity hereafter we call it "surface" and the rate of change of $n$ can be expressed as follows:

\begin{equation}
\frac{dn(t)}{dt}=-n(t)\gamma_{des}(t)+(n_{max}-n(t))\gamma_{abs}(t)
\label{eq:main}
\end{equation}

the first term $-n(t)\gamma_{des}(t)$ represents the molecules that leave the surface per unity of time. The parameter $\gamma_{des}(t)$ is the probability that each of the $n$ molecules is desorbed and it is a function of time, as we will see below.

The other addendum $(n_{max}-n(t))\gamma_{abs}(t)$ represents the rate of absorption, that is, in the same fashion, given by the product of a probability and a number of particles. We define $n_{max}$ as the number of particles that realize a complete coverage on the surface, thus $(n_{max}-n(t))$ represents how much space is available on the surface for absorption, measured in particle units. The probability to have a random gas particle landing on the surface is proportional to this space.

As mentioned above, these probabilities $\gamma_{des}$ and $\gamma_{abs}$ may not be constant, but depend on, for example, time variable gas pressure, accordingly to a linear relation:
\begin{eqnarray}
\gamma_{abs}&=\alpha_{abs}(g^{abs}_{mx}-g_{mn})P+g_{mx} \\
\gamma_{des}&=\alpha_{des}(g^{des}_{mn}-g_{mn})P+g_{mx} \nonumber
\label{eq:gamma}
\end{eqnarray}
where $g_{mx}$ and $g_{mn}$ are two offsets that represent the range of pressure and $\alpha$ is the coefficient that fit the experimental data.
In the differential equation (\ref{eq:main}), the first term on the right is clearly an exponential decay, whereas the second shows a logistic nature, so the three equations (\ref{eq:prop}), (\ref{eq:main}) and (\ref{eq:gamma}) altogether result in an exponential growth or exponential decay behavior, depending on the pressure parameter. 

\begin{figure}[!ht]
\begin{center}\leavevmode
\includegraphics[width=8.5cm]{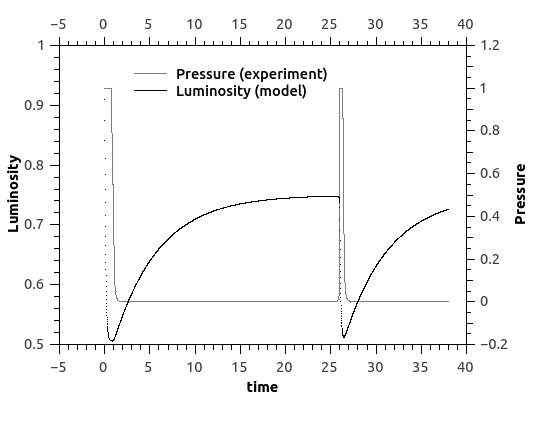}
\caption{A simulation of $n_{max}=1000$ molecules for the model described in equation (\ref{eq:main}) and equation (\ref{eq:gamma}). The continue curve represent the pressure obtained from experimental data, and the dotted profile is the simulation of the luminosity. We integrated using an Eulero method with time step $dt=0.01$ seconds, $\gamma_{abs}$ and $\gamma_{des}$ were set to 3 and 0.1 respectively. Other parameters in equations (\ref{eq:prop}), (\ref{eq:main}) and (\ref{eq:gamma}) were $I_{max}=1$, $\beta=5*10^{-4}$, $g^{abs}_{mx}=3$, $g^{des}_{mx}=0.1$ and $g_{mn}=0.1$. Vertical axis for luminosity is in arbitrary units, for pressure is Atmosphere. Horizontal axis is time in seconds. The pressure values are real data taken from the experiment.
}
\label{fig:simPress}
\end{center}
\end{figure}

If we integrate numerically the differential equation (\ref{eq:main}) and run a simulation with the experimental variable pressure data inserted in equation (\ref{eq:gamma}) we obtain the result in figure \ref{fig:simPress} that is strikingly similar to the behavior found in the experiments suggesting that is a dynamically changing coverage of spurious gas molecules absorbed on the surface that causes the photoluminescence variations observed experimentally.

We found that photoluminescence has lower intensity and shows unevenly distributed intense luminous centers\cite{mich:2004} (bright spots) and some blinking points when the level of absorption coverage is high on the sample surface. In contrast, when vacuum is drawn and spurious gas particles are desorbed, sample surface emission has higher intensity and it is homogeneous, with almost no bright spots or blinking points.

We speculate that unstable blinking and the so called optical memory effect are caused by absorbed material on the surface of the sample. In particular, the presence of O$_2$ in the proximity of the surface, induce adsorption that influence the excitation of GaN layers, not not only the InGaN layer.

We developed a working phenomenological model that fits properly the experimental curves that relate the ambient pressure and the sample photoluminescence.

This study wants to contribute to the understanding of fundamental processes involved in the  emission phenomena of InGaN materials and to help break ground to improve the efficiency and reliability of next generation nitride semiconductor devices.

\begin{acknowledgments}
This study was partially supported by the Japanese grant  KAKEN (project number 24560014) and by the Yokohama City University Sabbatical fund 2015-2016.
\end{acknowledgments}
% Produces the bibliography via BibTeX.

\bibliography{biblDatabase_all_e} 

\newpage

%\begin{figure}[!ht]
%\begin{center}\leavevmode
%\includegraphics[width=8.5cm]{Graph8.jpg}
%\caption{Profile of the luminosity exponential decay when pressure returns to 1 Atmosphere after a period of high vacuum state. The experiment is performed in the same conditions as in figure \ref{fig:photoeff}. The continous curve is a single exponential decay fit.}%
%\label{fig:airPr}
%\end{center}
%\end{figure}

\end{document}